\documentclass[useAMS,usenatbib]{mn2e}

\usepackage{graphicx}
\usepackage{amsfonts,amssymb}
\usepackage{times}

\def\rot{\mathop{\rm rot}\nolimits}
\def\div{\mathop{\rm div}\nolimits} 
\def\gsim{\lower.4ex\hbox{$\;\buildrel >\over{\scriptstyle\sim}\;$}} 
\def\lsim{\lower.4ex\hbox{$\;\buildrel <\over{\scriptstyle\sim}\;$}} 

\def\q{\qquad}

\def\beg{\begin{eqnarray}}
\def\ende{\end{eqnarray}}

\renewcommand{\Re}{\mathrm{Re}}
\newcommand{\Rm}{\mathrm{Rm}}

\newcommand{\Rin}{R_\mathrm{in}}
\newcommand{\Rout}{R_\mathrm{out}}
\newcommand{\etah}{\hat{\eta}}
\newcommand{\Pm}{\mathrm{Pm}}
\newcommand{\Ha}{\mathrm{Ha}}

\newcommand{\Om}{{\it \Omega}}
\renewcommand{\vec}[1]{\mbox{\boldmath $#1$}}
\newcommand{\mperm}{\mu_0}
\newcommand{\mdiff}{\eta}

\title[Eddy viscosity and turbulent Schmidt number by magnetic instability]
      {Eddy viscosity and turbulent Schmidt number by kink-type instability of strong toroidal magnetic fields}

\author[G. R\"udiger, M. Gellert and M. Schultz]
       {G\"unther R\"udiger\thanks{E-mail: gruediger@aip.de},   
        Marcus Gellert and Manfred Schultz\\ 
        Astrophysikalisches Institut Potsdam, An der Sternwarte 16, D-14482 Potsdam, Germany} 

\begin{document}

\pagerange{\pageref{firstpage}--\pageref{lastpage}} \pubyear{2008}
\maketitle
\label{firstpage}

\begin{abstract}
The potential of the nonaxisymmetric magnetic instability to transport angular momentum and to mix chemicals is probed considering  the  stability of a  nearly uniform 
 toroidal field between conducting cylinders with different rotation rates. The fluid between 
the cylinders is assumed as incompressible and to be of  uniform density. With a linear  theory the neutral-stability maps for  $m=1$ are 
computed. Rigid rotation must be subAlfv\'enic to allow instability while for differential rotation with negative shear 
also an unstable domain with superAlfv\'enic rotation exists. The 
rotational quenching  of the magnetic instability is   strongest for magnetic Prandtl number $\rm Pm =1$  and becomes much  weaker for $\Pm\neq 1$.
  
The effective angular momentum transport by the instability is directed outwards(inwards)  for subrotation(superrotation). 
The resulting magnetic-induced eddy viscosities  exceed the microscopic 
values  by factors of 10--100. This is  only true  for  superAlfv\'enic flows; in the strong-field limit the values remain much smaller.

The same instability also quenches concentration gradients of chemicals by  its nonmagnetic  fluctuations. The corresponding diffusion coefficient  remains always smaller than the magnetic-generated eddy viscosity.  A Schmidt number of order 30 
is found as the ratio of the effective viscosity and the diffusion coefficient. The magnetic instability transports much 
more angular momentum than that it mixes chemicals.
\end{abstract}

\begin{keywords}
instabilities --  magnetic fields  -- angular momentum transport -- turbulent diffusion.
\end{keywords}

\section{Introduction}
The current-driven  magnetic   instability is an important  instability of toroidal fields  which is basically 
nonaxisymmetric  (`kink-type') \citep{tayler57,tayler73,vanda}. In real fluids a toroidal field becomes unstable at a certain magnetic field amplitude 
depending on the radial profile of the field. A global solid-body rotation of the system influences 
the critical field amplitude \citep{acheson78, pitts85}. It is strongly increased for fast rotation so that we shall speak about 
a `rotational  quenching' of the Tayler instability (TI). 

\begin{figure}
    \center \includegraphics[height=8cm,width=8.0cm]{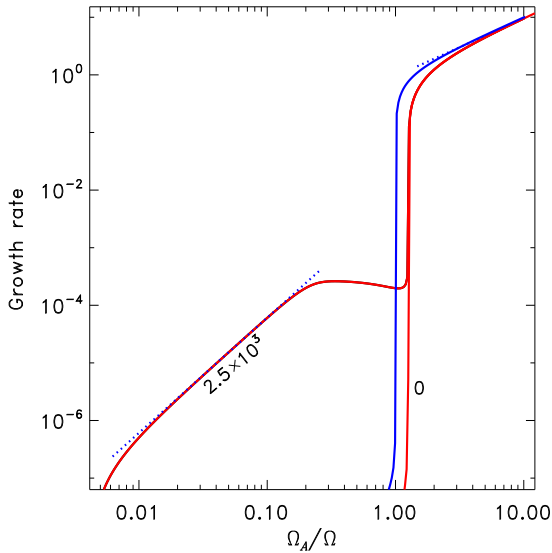}
    \caption{The growth rates normalized with the rotation rate for the kink-type instability of an equatorsymmetric toroidal 
             field in a sphere, here with $\Pm=5\cdot 10^{-3}$, (see Kitchatinov \& R\"udiger 2008). The two red lines are for 
             $\rm q=0$ (right) and the solar value $\rm q=2500$ (left). The blue line gives the limits for both $g=0$ and $\chi\to \infty$.}
    \label{onebelt}
\end{figure}

It has been shown by \citet{kit} that this  magnetic instability in rotating spheres with an 
equatorsymmetric magnetic field (peaking at the equator) strongly differs for    subAlfv\'enic  rotation  ($\Om< \Om_{\rm A}$) and  superAlfv\'enic 
rotation ($\Om> \Om_{\rm A}$). Here $\Om_{\rm A}$ is the Alfv\'en frequency
\beg
\Om_{\rm A}=\frac{B_\phi}{\sqrt{\mu_0\rho} R^\ast},
\label{Alf}
\ende
with $\Om$ the rotation rate and $R^\ast$  a characteristic radius. The growth rates of the instability are many 
orders of magnitudes smaller for weak fields than for strong fields (Fig.~\ref{onebelt}). 
For $\Om_{\rm A} > \Om$ the growth of the instability is very fast, its growth rate  is of the order of the 
rotation rate. The values of the microscopic dissipation  parameters (viscosity, thermal diffusion and magnetic resistivity) 
have a very small influence on the stability behavior of  strong fields. The growth rate runs linearly  with the magnetic field amplitude. 

The situation strongly differs  for  superAlfv\'enic flow  with $\Om_{\rm A}<\Om$ where instability  only exists for finite values 
of the heat conductivity. It does not exist for both $\chi=0$ and $\chi\to \infty$.  The latter is obvious as for extremely  
fast thermal diffusion the density fluctuations become zero and the buoyancy term disappears. The adiabatic case with 
$\chi=0$ is also almost without density fluctuations so that no extra instability appears \citep{acheson78}.

It seems as would the lower limit of the magnetic field for neutral instability be fixed by the Roberts number
\beg
{\rm q}=\frac{\chi}{\eta}.
\label{roberts}
\ende
There is a  minimum of the field amplitude sufficient for instability which is well represented in Fig.~\ref{onebelt} by 
the curve for $\rm q=2500$. Note, however, the smallness of the corresponding growth rates. The maximum growth rate in 
the weak-field domain is $\simeq 10^{-4} \Om$, i.e. the growth time is  more than 1000 rotation times. It is an open and challanging 
question whether such slow nonaxisymmetric instabilities are important for the dynamics of the stellar interior. 

The instability for strong fields with (${\Om}_{\rm A} \geq {\Om}$) is much faster and it grows linearly with the magnetic 
field independent of the rotation rate. The instability does not `feel' both the thermal diffusion and  the density 
stratification. With nonlinear simulations we shall answer the question how strong the angular momentum transport and 
the turbulent mixing due to the Tayler instability are. The complex destabilizing influence of finite thermal diffusion 
(see Fig. 1 in Acheson's paper) is ignored in the calculations what makes a drastic simplification for the theory. 
Also for reasons of simplicity the computations are done in cylindric Taylor-Couette geometry where the differential 
rotation can be fixed by choice of the  rotation rates of the inner and the outer cylinder.

For liquid sodium the Roberts number is small, $\rm q=10^{-3}$. The neglect of the thermal diffusion is thus 
fully realistic for MHD experiments in the laboratory. It is also realistic in stellar physics if the scenario of decaying 
fossil magnetic fields is considered. \citet{moss} finds poloidal fields of order 1000 G surviving in the inner radiation zones 
after the Hayashi track. Only after a short time, differential rotation produces the toroidal field component which tends to 
dominate the poloidal fields. The resulting Maxwell stress suppresses the differential rotation after the Alfv\'en travel time of 
the poloidal field unless the toroidal field becomes unstable. It seems to be clear that in order to be effective, the growth rate 
of the instability must be of order of the rotation rate rather than orders of magnitudes smaller.

The resulting instabilities form a nonaxisymmetric  pattern of flow and field which  transports both angular momentum 
and chemicals like lithium. We are interested in the possibility to express these fluxes by the gradients of mean quantities.  
The angular momentum flux for rigid rotation in our simplified  model proves to be very small. It is thus reasonable to express the 
angular momentum transport by an eddy viscosity $\nu_{\rm T}$ and also to express the flux of chemicals by an eddy diffusivity 
$D_{\rm T}$. The amplitudes of the transport parameters are computed by nonlinear numerical simulations.

\section{Astrophysical motivations}
There is a variety of astrophysical  applications. Only some  of them are discussed in the present paper.
\subsection{Rigid rotation in the solar interior}
We know from the results of the helioseismology that  the solar radiative interior  rotates  rigidly. This  is insofar surprising  
as the microscopic viscosity  with $\nu\lsim 10\ {\rm cm}^2/{\rm s}$ of the hot  solar plasma yields  a diffusion time of more than $10^{13}$ 
yr so that  any initial rotation law (which  certainly  existed) should be  still  existing without any other influence. One needs effective 
viscosities of $10^5\ {\rm cm}^2/{\rm s}$ to explain the decay of an initial rotation law  within the lifetime of the Sun 
(see also \citealt{yang}). Future results of asteroseismology will show how general this statement is. Also the inclusion of the 
angular momentum transport by a poloidal magnetic field subject to differential rotation needs for the explanation of the solar spin-down 
the increase of the microscopic viscosity by a few orders of magnitude \citep{rued96}. We have to test whether the angular momentum transport 
by magnetic instabilities of  fossil magnetic fields is strong enough to produce the quasi-solid inner rotation of the Sun.

\subsection{Diffusion of chemical elements}
The lithium at the surface of cool MS stars decays with a timescale of 1 Gyr. It is burned at temperatures in excess of $2.6 \times 10^6$ K 
which exists about 40.000 km below the base of the convection zone. There must be a diffusion process  through this layer to the location of 
the burning temperature. Hence, we have to look for a slow process which, however, is only one or two orders of magnitude faster than the 
molecular diffusion. The molecular diffusion at the bottom of the solar convection zone results as $D_{\rm micro}\simeq 30$\ cm$^2$/s 
\citep{barnes} which must be increased to about $10^3$ cm$^2$/s as the consequence of a hypothetical instability. Note the smallness of this 
quantity: the  plasma velocities for such a diffusion coefficient are by many orders of magnitude smaller than the convective velocities of 
about 100 m/s at the bottom of the convection zone.

There is thus a good motivation to look for a mechanism with a diffusion coefficient $D_{\rm T} > D_{\rm micro}$ but forming
a turbulent Schmidt number with 
\beg
{\rm Sc}= \frac{\nu_{\rm T}}{D_{\rm T}}\gg 1
\label{Pd}
\ende
(see \citealt{zahn}). The expressions formulated by \citet{eggenberger} on the basis of the `Tayler-Spruit' dynamo scenario lead for a toroidal 
magnetic field of about 100 Gauss to $\rm Sc \simeq 100$ (where the turbulent magnetic diffusivity has been used as the value for $D_{\rm T}$). 
This is a reasonable value for the Schmidt number. It is doubtful, however, that a toroidal field of 100 Gauss is unstable. For the upper part 
of the solar radiative  core we found 600 Gauss as the minimum amplitude of a toroidal field to become unstable under the presence of the 
actual rotation rate, density stratification and diffusion processes \citep{kit}. A dynamo action could not be confirmed so far 
\citep{brun,gellert08}.

Considering the transport processes in the radiative interior of massive (15 $M_\odot$) MS stars  on the basis of analytic  expressions 
by \citet{spruit}, \citet{maed03,maed05} computed viscosities up to $10^{13}\ {\rm cm}^2/{\rm s}$ for a rotational velocity of 300 km/s. 
The resulting diffusion time is only a few 100 yr. An extreme chemical mixing is the immediate consequence. With similar expressions \citet{yoon}
present evolutionary models of rotating stars with different metalicities.

Angular momentum transporters could be Maxwell stress by both poloidal fields ($B_R$) and toroidal fields ($B_\phi$) which are connected 
by the differential rotation. If the Maxwell stress can be expressed by an effective viscosity and a very similar expression is used 
for chemical diffusion processes then the material mixing becomes so strong that the resulting evolution models are incompatible with 
observations. \citet{brott} demonstrate that the corresponding chemical mixing in massive stars must even be neglected to understand 
the observations.  

Note that the material mixing only results by the action of the kinetic part of the momentum transport tensor rather than by 
its magnetic part so that the diffusion tensor can be approximated by
\beg
D_{ij}\simeq \frac{1}{2} \tau_{\rm corr}\ \langle u_i'(\vec{x},t)u_j'(\vec{x},t)\rangle 
\label{Dij}
\ende
(for details see \citealt{rued01}). There is no magnetic influence in the diffusion equation except the magnetic suppression of the 
correlation tensor of the fluctuations. 

If the correlation time of the instability scales with its growth time then the coefficient of the diffusion of chemicals can simply  be written as
\beg
D_{\rm T}\simeq  \frac{{\langle u_R'^2\rangle}}{\gamma} .
\label{D}
\ende
For strong fields one can estimate the growth rate $\gamma\simeq \Om$ (see below). Inserting numbers of the solar  tachocline ($D_{\rm T} \simeq 10^3$\ cm$^2$/s, $\Om\simeq 10^{-6}$\ s$^{-1}$) 
leads to $u_R'$ of order 0.1 cm/s. It is not easy to produce such a  weak turbulence intensity in radial direction. The idea is that the 
stable density stratification below the convection zone suppresses the vertical turbulence component so that a highly anisotropic turbulence 
field evolves \citep{vinc,toq}. It has been shown, however, that due to the stellar rotation even a strictly horizontal flow pattern obtains 
radial components which are able to transport chemicals through the layer below the convection zone \citep{rued01}. The result is that the Li 
depletion becomes anticorrelated with the (rapid) stellar rotation or -- with other words -- the fast rotators should possess the highest surface 
concentrations of lithium. This consequence of the rotational quenching of the turbulence has indeed been found in the Pleiades 
cluster \citep{xing}. Such a rotation-lithium correlation, however, could also be the consequence of the rotational quenching of the TI 
which is considered in the present paper.

\section{Magnetic  Taylor-Couette flow}
\subsection{The model}
\begin{figure}
    \center \includegraphics[width=6.0cm]{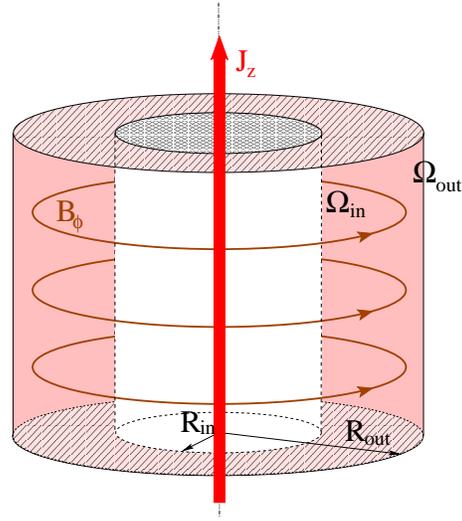}
    \caption{Two concentric cylinders with radii $\Rin$  and  $\Rout$  rotate with $\Om_{\rm in}$ and
	     $\Om_{\rm out}$. $B_\phi$ is the given toroidal magnetic field produced by  axial  currents inside and outside the inner cylinder.}
    \label{geom}
\end{figure}

A Taylor-Couette container is considered  confining  a  toroidal magnetic field with given amplitude at the cylinders  rotating with different 
rotation rates  (see Fig. \ref{geom}). In order to simulate the situation at the bottom of the convection zone (or even at its top) the gap 
between the cylinders is considered as small. The outer radius is $\Rout$ and the inner radius is  $\Rin$.
The fluid confined between the cylinders is assumed to be incompressible with uniform density and dissipative with  kinematic viscosity $\nu$ and  magnetic diffusivity
$\mdiff$. Derived from  the conservation law of angular momentum the angular velocity of the rotation is
\beg
 \Om=a+\frac{b}{R^2}
\label{1}
\ende
with
\beg
 a=\frac{\mu_\Omega-\etah^2}{1-\etah^2} \Om_{\rm in}, \q 
 b=\frac{1-\mu_\Omega}{1-\etah^2}\Rin^2 \Om_{\rm in},
\ende
where
\beg
 \hat\eta=\frac{R_{\rm{in}}}{R_{\rm{out}}}, \q
 \mu_\Omega=\frac{\Om_{\rm{out}}}{\Om_{\rm{in}}}.
 \label{mu}
\ende
$\Om_{\rm in}$ and $\Om_{\rm out}$ are the imposed rotation rates of the inner and outer cylinders.
After the Rayleigh stability criterion the flow is hydrodynamically stable for $\mu_\Omega>\etah^2$, i.e. $\mu_\Omega>0.25$ for  $\etah=0.5$.
We are only interested in hydrodynamically stable regimes so that $\mu_\Omega>\etah^2$ should always be fulfilled. Rotation laws with 
${\rm d}\Om/{\rm d}R<0$  are described by $\mu_\Omega<1$; $\mu_\Omega=1$ gives rigid rotation.
 
Also the magnetic profiles are restricted. The solution of the stationary induction equation without inducing flow reads 
\beg
B_\phi=A R+\frac{B}{R}.
\label{basic}
\ende
(in cylindric geometry. $A$ and $B$ are the fundamental quantities; $A\cdot R$ 
corresponds to a uniform axial current ($I=2A/\mu_0$) everywhere within $R<R_{\rm out}$, and $B/R$ corresponds to an additional current 
only within $R<R_{\rm in}$. The fields with $A=0$ are called the {\em current-free} solutions (current-free between the cylinders). Both terms in (\ref{basic}) are responsible for their own instability  where one of them only exists in combination with differential rotation.

 In analogy with $\mu_\Omega$ it is useful to define the quantity
\beg
\mu_B=\frac{B_{\rm{out}}}{B_{\rm{in}}},
\ende
measuring the variation in $B_\phi$ across the gap. Here we shall consider mainly fields which are rather uniform in radius, i.e.  
$\mu_B \simeq 1$. It is $B=R_{\rm in} R_{\rm out} A$ for this choice so that both parts of (\ref{basic}) are involved. The condition 
$A=0$ leads to profiles with $\mu_B=0.5$,  and the condition $B=0$ leads to $\mu_B=2$ both for our standard case  $\hat\eta=0.5$. After the stability criterion for toroidal fields against axisymmetric perturbations one 
finds stability of profiles with  $0\leq \mu_B\ \leq 1/\hat \eta$  (see Shalybkov 2006). The magnetic fields discussed in the present  paper are thus stable against disturbances with $m=0$ (see below).

In the following we shall fix  $\mu_B=1$  but  the rotation law and the magnetic Prandtl number
\beg
\Pm=\frac{\nu}{\eta}.
\label{pm}
\ende
are varied. If the results shall be applied to stellar  convection zones  or e.g. to the fields in galaxies 
the magnetic Prandtl number must be replaced by its  turbulent value.

\subsection{The  equations}
The dimensionless incompressible MHD equations are
\begin{eqnarray}
 \lefteqn{   {\rm Re}    \frac{\partial\vec{u}}{\partial t} +{\rm Re} (\vec{u} \cdot \nabla)\vec{u} =
               -\nabla P + \Delta \vec{u} + {\Ha^2} \rot \vec{B} \times \vec{B},} \nonumber\\
 \lefteqn{   {\rm Rm}    \frac{\partial \vec{B}}{\partial t} = \Delta \vec{B} + {\rm Rm} \rot (\vec{u} \times \vec{B}),} 
\end{eqnarray}
with $\div{\vec{u}} =  \div{\vec{B}} = 0$ and with the Hartmann number
\beg
    \Ha = \frac{B_{\rm in} R_0}{\sqrt{\mperm \rho \nu \mdiff}}.
\ende
$R_0=\sqrt{\Rin(\Rout - \Rin)}$ is used as the unit of length, $\eta/R_0$ as the unit of velocity and $B_{\rm in}$ as the unit of the magnetic fields. 
Frequencies including the  angular velocity $\Om$ of the rotation are normalized with the  rotation rate $\Om_{\rm in}$. The Reynolds number $\Re$ is defined as 
\beg
\Re=\frac{\Om_{\rm in}  R_0^2}{\nu},
\label{Rey}
\ende
and the magnetic Reynolds number as $\Rm=\Pm \cdot \Re$. The Lundquist number is simply $\rm S= \Ha \sqrt{\Pm}$. 

The boundaries are assumed as impenetrable and stress-free and as perfect conductors. 
 
The solutions of the linear equations are free to an arbitrary real parameter of any sign. We  thus cannot know the sign of the flow and/or the field. 
However, for quadratic expressions such as the correlation tensor or the electromotive force we can compute the sign as all the solutions are multiplied 
with one and the same parameter.

Let us apply this idea to the angular momentum transport
\begin{equation}
 T_R=\langle u_R' u_\phi' - \frac{1}{\mu_0 \rho} B_R' B_\phi'\rangle .
\label{T}
\end{equation}
The averaging procedure is thought  to be  an integration over the azimuth $\phi$. The question  we  shall answer is whether $T_R>0$ for ${\rm d}\Om/{\rm d}R<0$ 
and $T_R<0$ for ${\rm d}\Om/{\rm d}R>0$. If this is true then the angular momentum flows towards the minimum of the angular velocity and one can introduce an 
eddy viscosity $\nu_{\rm T}$, i.e.
\begin{equation}
 T_R=- \nu_{\rm T} R \frac{{\rm d}\Om}{{\rm d}R}
\label{T1}
\end{equation}
with positive $\nu_{\rm T}$. The sign  of $\nu_{\rm T}$ can be computed  with the linear  theory. 
We have computed  this quantity with both linear and nonlinear approximations. With the linear theory  bifurcation diagrams for various rotation 
laws and magnetic Prandtl numbers are established and it is shown that the angular momentum (kinetic plus magnetic) flows opposite to the gradient 
of the angular velocity. It is thus possible in this case to define via (\ref{T1})  an eddy viscosity on the basis of the Tayler instability. With the nonlinear 
simulations the amplitude of this eddy viscosity is computed. 

With $U=\langle u_R'^2\rangle$ the ratio of $\nu_{\rm T}$ and $D_{\rm T}$ results to
\beg
 {\rm Sc}=\frac{1}{q} \frac{T_R}{U}
\label{nutD}
\ende
with $q=-{\rm d} {\rm log} \Om/{\rm d} {\rm log} R$. Without Maxwell stress this quantity is certainly of order unity but it can acquire much higher 
values if the magnetic-induced angular momentum transport dominates.

 \section{Marginal instability}
In the linear approximation the solutions can be found in form of normal modes in accord to 
\beg
 F=F(R) {\rm e}^{{\rm i}(kz+m\phi+\omega t)}.
\label{four}
\ende
Here $m$ is the azimuthal quantum number of the resulting cell pattern. For instabilities the growth rate $\gamma= - \Im(\omega)$ must be positive 
and the neutral lines are defined by $\gamma=0$. In this Section the neutral lines for the magnetic  instability with $m=1$ are presented. The wave number $k$ is optimized so that the resulting Reynolds number becomes a minimum for given Hartmann number.
 \begin{figure}
   \includegraphics[width=8.0cm,height=5.0cm]{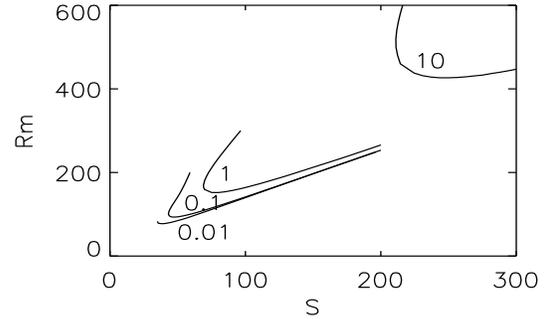}
   \caption{The lines of neutral instability for $A=0$ (AMRI). The curves are marked with their magnetic Prandtl numbers. The toroidal fields are 
            current-free between the (conducting) cylinders. It is $\mu_B=\etah=0.5$, $\mu_\Omega=0.5$.}
   \label{map1}
\end{figure}
\subsection{The maps of neutral instability}
Let us start with the phenomenon that even {\em current-free} toroidal fields can become unstable if they are subject to differential rotation. For 
$\etah=0.5$ follows $\mu_B=0.5$ for the magnetic profile, and  the differential rotation may be as weak as $\mu_\Omega=0.5$ simulating a (galactic) rotation law $\Om \propto 1/R$. The map for  neutral instability of this interaction 
for various magnetic Prandtl numbers is given in Fig. \ref{map1}. The disturbances with $m=1$ can grow although the magnetic field or the 
differential rotation alone are stable. We have called this phenomenon as Azimuthal Magnetorotational Instability (AMRI, see \citealt{rued07}, also Ogilvie \& Pringle 1996). 
It is very characteristic for AMRI that it disappears for rigid rotation but also  if the (differential) rotation becomes too strong. This is the reason 
why both lines of the marginal stability have the same positive slope.    

\begin{figure}
   \includegraphics[width=8.0cm,height=5.0cm]{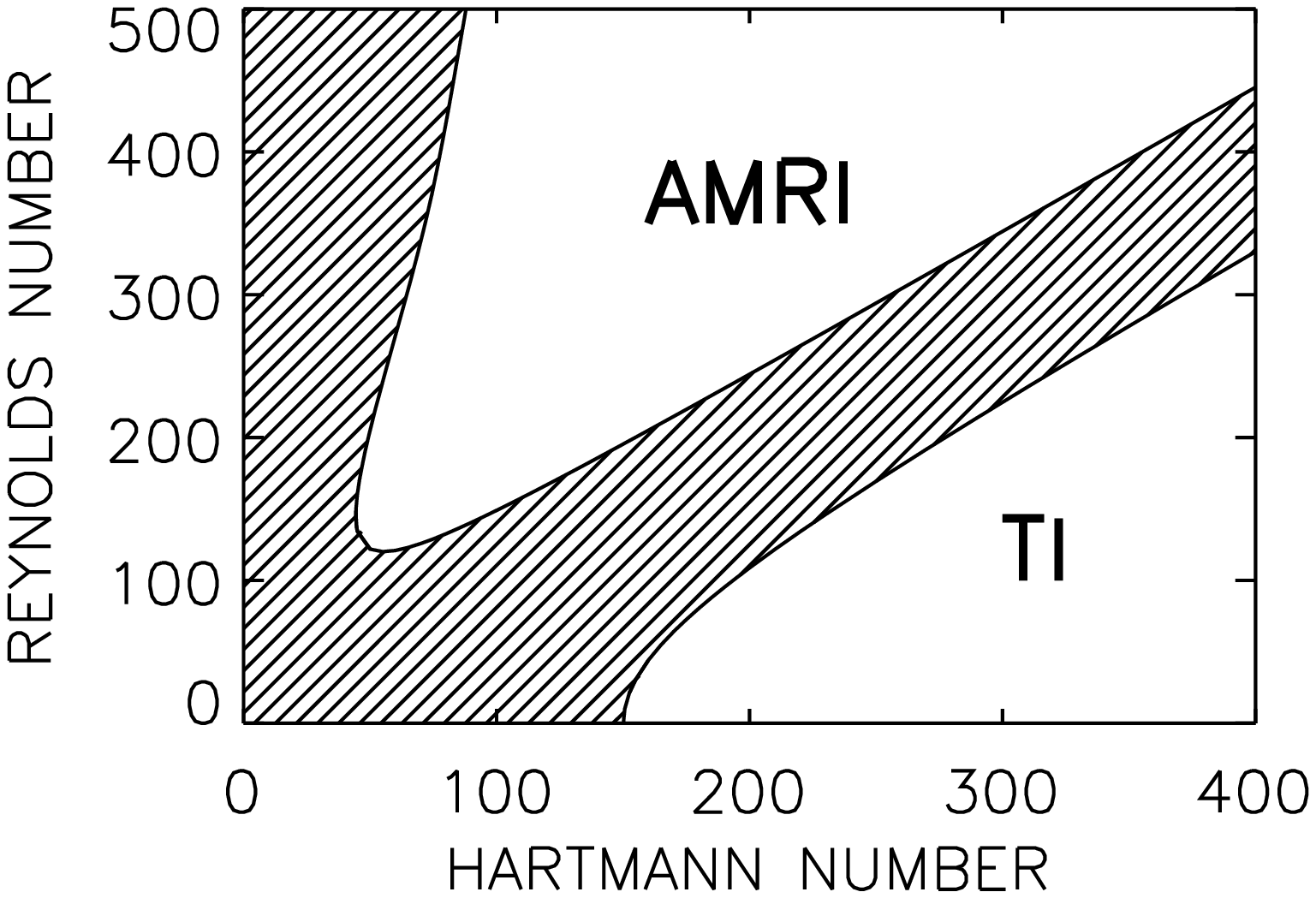}
    \includegraphics[width=8.0cm,height=5.0cm]{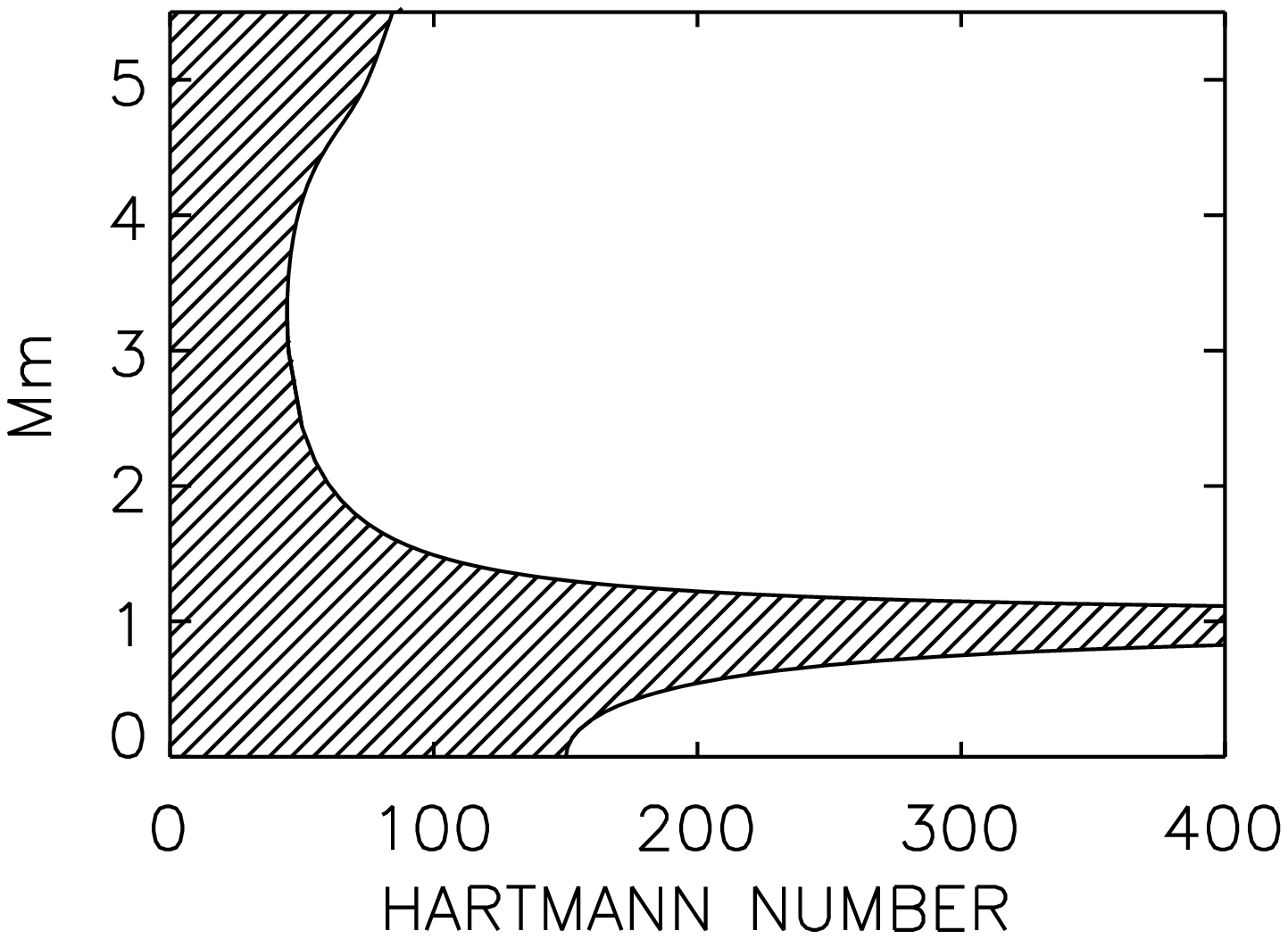}
   \caption{The marginal-instability map for toroidal fields with $A\simeq B$ ($\mu_B=1$). Top: Rotation vs magnetic field. The AMRI domain and the TI domain are separated by a stable branch.  Bottom:  The  stable branch  between  the AMRI domain and the TI domain  corresponds to  the line $\rm Mm=1$. 
        It is $\Pm=1$, $\mu_\Omega=0.5$, $\etah=0.5$. }
   \label{map2}
\end{figure}

An  axial current ($I= 2A$) is not necessary  for   AMRI. But the  AMRI  survives the addition  of axial electric currents in the fluid conductor (i.e. $A\neq 0$). It also appears in the instability maps of toroidal fields with 
electric currents between the cylinders (Fig. \ref{map2}). In this figure the results for $\Pm=1$ and for weak subrotation ($\mu_\Omega=0.5$) 
are given for almost homogeneous fields with $\mu_B=1$. For this magnetic profile both the coefficients $A$ and $B$ in Eq. (\ref{basic}) are finite. The characteristic AMRI domain only vanishes for magnetic profiles with $B=0$.

Without  rotation  the magnetic field alone is unstable for $\rm Ha\geq 150$ because of the action of the 
axial current. We shall call this domain of subAlfv\'enic rotation  as the domain of the Tayler instability (TI). It also 
exists for $B=0$. The critical Hartmann number for vanishing rotation does  not depend on Pm. In Fig. \ref{map2}  two unstable 
domains with different growth rates exist. The growth rate is of order $\Om$ in the TI domain but it is  smaller in the AMRI domain (see below). Both domains 
are separated by a stability branch along the line $\rm Rm=Ha$ which disappears, however,  for $\rm Pm>1$ (see \citealt{rued07}). The AMRI domain in Fig. \ref{map2}, of course,  disappears for 
solid-body rotation.
\begin{figure}
   \includegraphics[width=8.0cm,height=5.0cm]{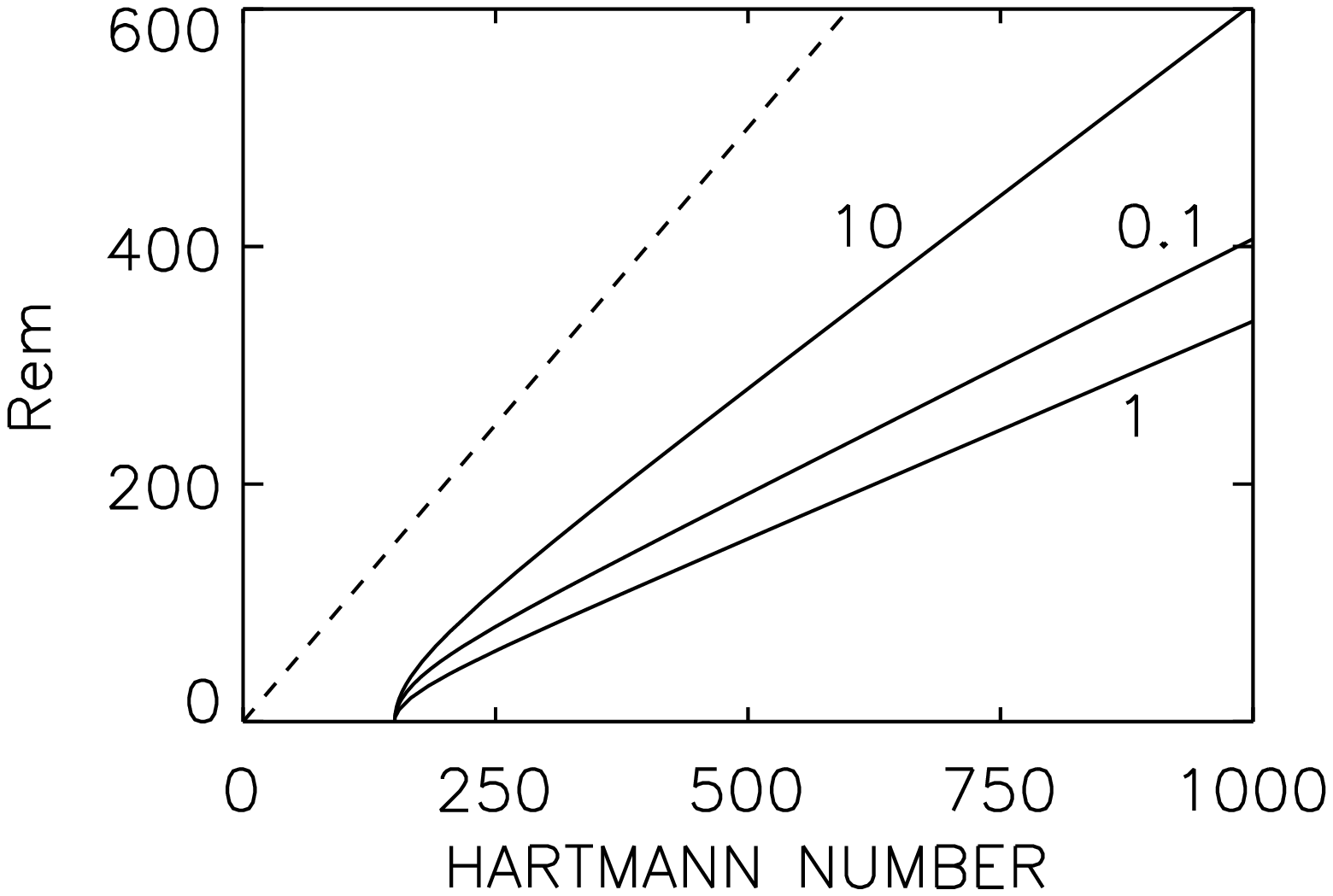}
   \includegraphics[width=8.0cm,height=5.0cm]{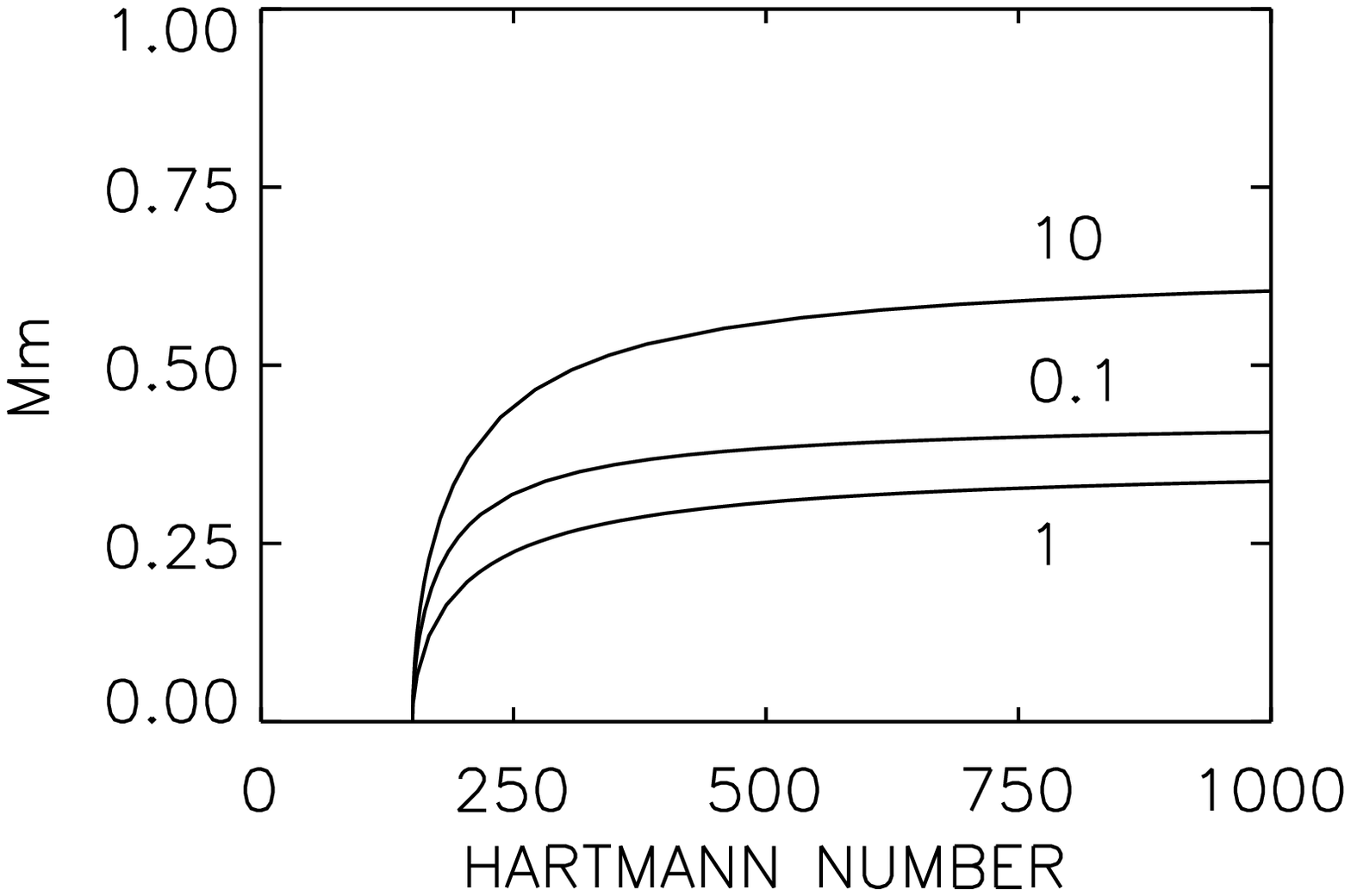}
   \caption{Top: The rotational quenching of the TI for rigid rotation and for various Pm. The dashed line gives $\rm Mm=1$. Note that 
            for $\rm Pm=1$ the rotational quenching of the magnetic instability is strongest. Bottom: TI is a 
            subAlfv\'enic phenomenon. $\etah=0.5$, $\mu_B=1$.}
   \label{map4}
\end{figure}

Note that the rotation basically {\em stabilizes} the instability \citep{pitts85}. The faster the rotation the stronger fields remain stable. 
We must keep this effect in mind for stellar applications. Magnetic fields which are stable for fast rotation become unstable during the 
stellar evolution with its continuous spin-down. Also the magnetic Prandtl number influences the rotational quenching. One finds the main result 
already by comparing the instability maps for rigid rotation only. The magnetic Prandtl number has been varied by two orders of magnitudes in 
Fig. \ref{map4}. It makes sense to interprete the results by means of the `mixed' Reynolds number
\beg
 {\rm Rem}=  \sqrt{{\Re \cdot \Rm}},
 \label{rem}
\ende
which by definition is symmetric in $\nu$ and $\eta$ as it is the Hartmann number. All fluids with the same product $\nu\cdot \eta$ have the same 
Rem and Ha. 
Consequently, the ratio of the mixed Reynolds number Rem and the Hartmann number Ha is a magnetic Mach number
\beg
 {\rm Mm}=  \frac{\rm Rem}{\rm Ha},
 \label{Mm}
\ende
in which the numerical values of viscosity and magnetic diffusivity do no longer  appear. $\rm Mm>1$ means superAlfv\'enic rotation and $\rm Mm<1$ describes subAlfv\'enic rotation.

\begin{figure}
   \includegraphics[width=8.0cm,height=5.0cm]{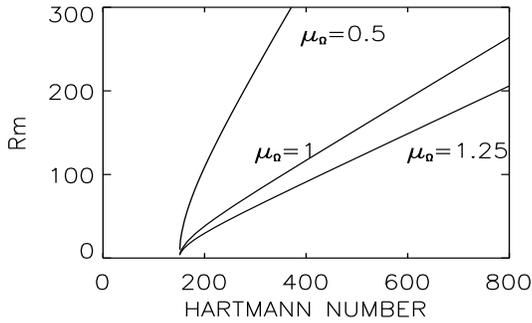}
   \caption{The marginal-instability map for subrotation ($\mu_\Omega=0.5$), for rigid rotation ($\mu_\Omega=1$) and for superrotation 
            ($\mu_\Omega=1.25$). The curve for $\mu_\Omega=0.5$ is taken from Fig. \ref{map2}. It is $\mu_B=1$, $\Pm=1$, $\etah=0.5$.}
   \label{map3}
\end{figure}

The rotational stabilization of the toroidal fields also depends on the form of the rotation law. It is weaker for subrotation and it is stronger for rigid rotation and for
superrotation (Fig. \ref{map3}). The trend is understandable as superrotation is the most stable rotation law in hydrodynamics. For given 
rotation rate the maximum field amplitudes which remain stable are much weaker for subrotation than for rigid rotation. With other words: 
rigid rotation stabilizes magnetic fields more effective than  subrotation. For very fast rotators, however,  
the nonaxisymmetric instability is always suppressed. For superAlfv\'enic rotation ($\Re \gg \Ha$) the magnetic fields become stable also  for rotation laws with negative shear.  

\subsection{Growth rates}
\begin{figure}
  \includegraphics[width=8.0cm,height=5.0cm]{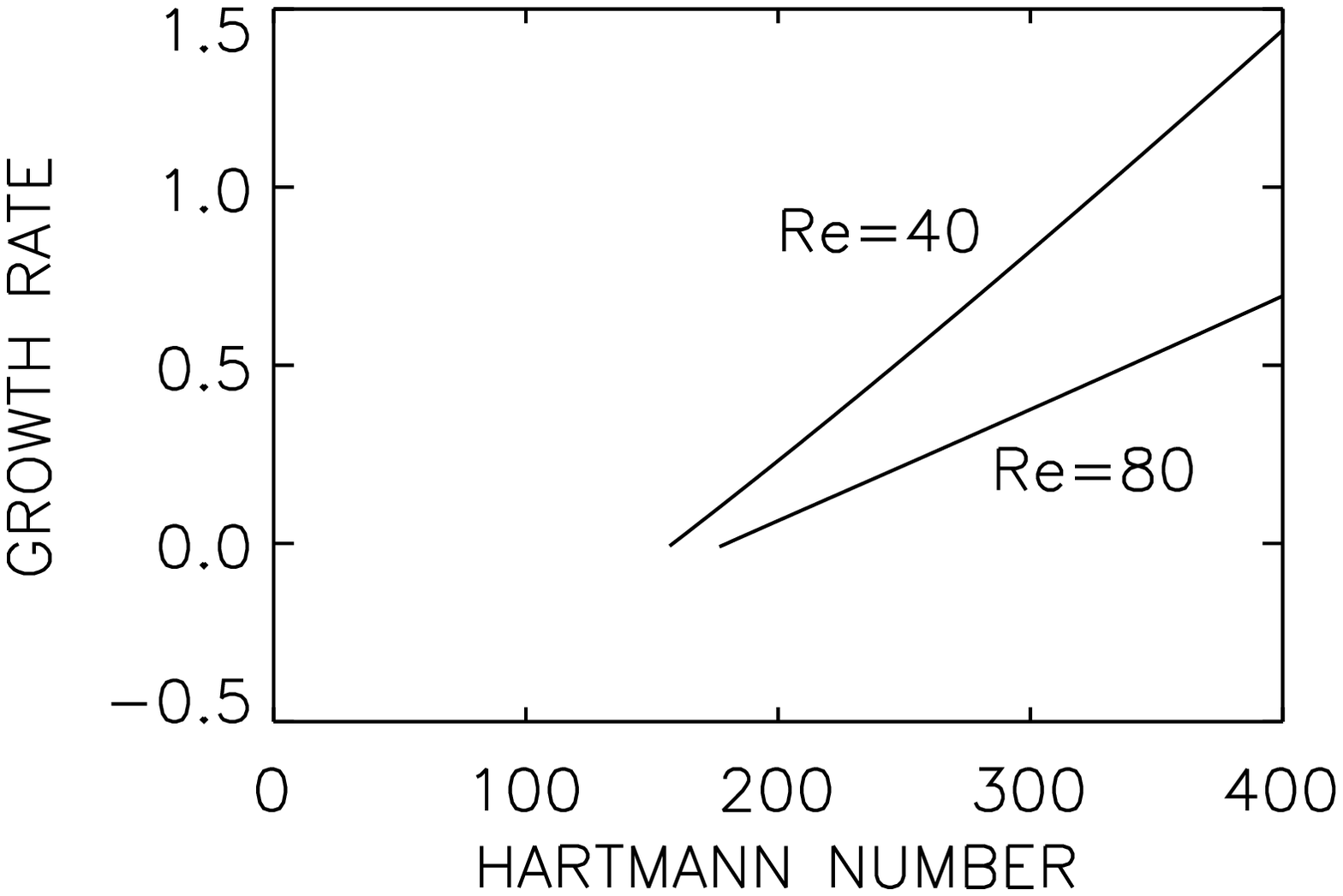}
   \includegraphics[width=8.0cm,height=5.0cm]{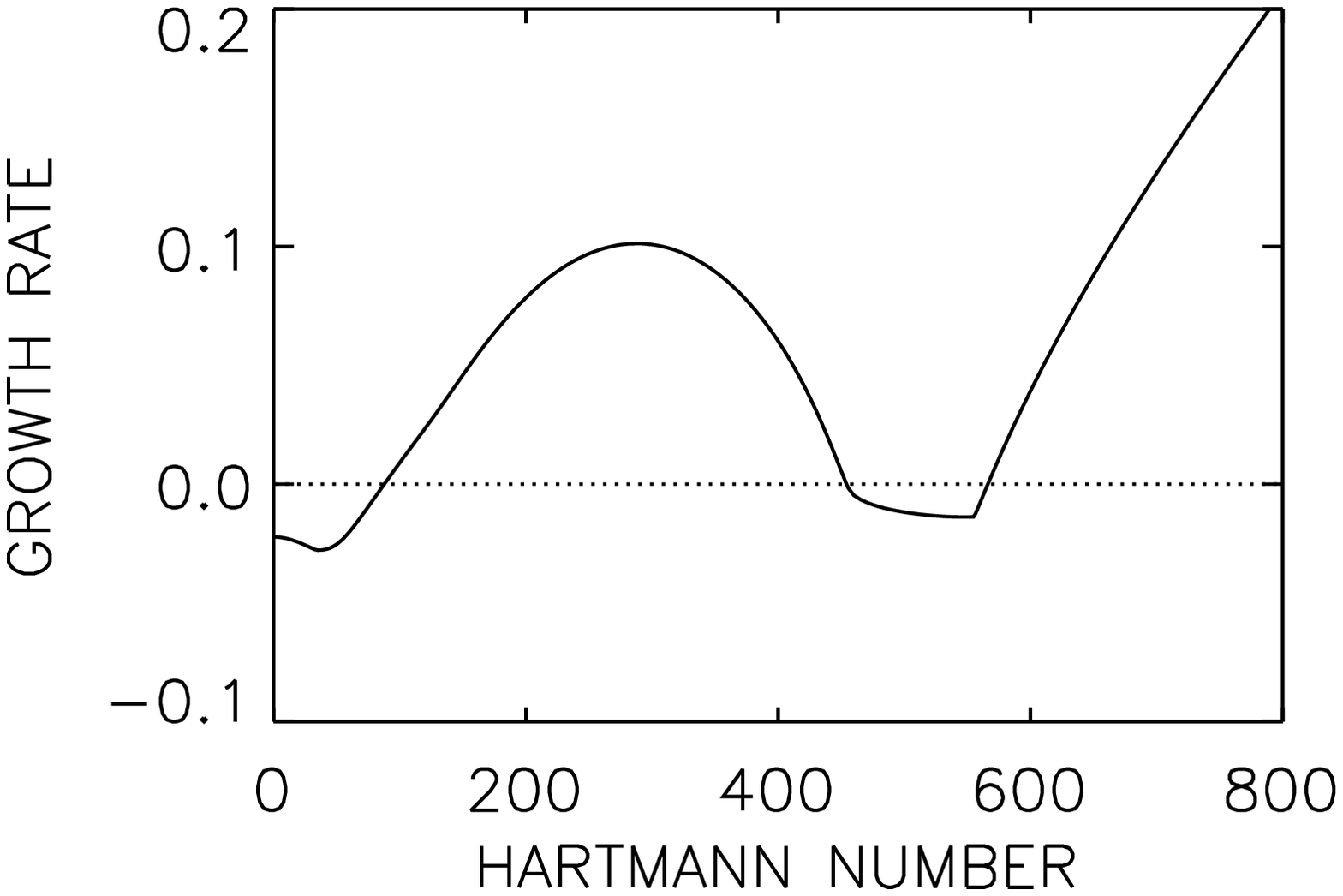}
  \caption{The (normalized) growth rates for  $\rm Pm=1$ (see Fig. \ref{map2}). Top: the TI domain. The curves are marked with their Reynolds numbers. 
           After multiplication with Re they are almost identical.  Bottom:  along the line $\rm Re=500$ of Fig. \ref{map2} (AMRI and TI).}
  \label{growth1}
\end{figure}

The growth rates $\gamma$ are the negative imaginary part of the frequency $\omega$ in Eq. (\ref{four}). They are given (normalized with the 
rotation frequency of the inner cylinder) in the Fig. \ref{growth1}. The growth time in units of the rotation time can be 
obtained as $\tau_{\rm growth}/\tau_{\rm rot}= 1/(2\pi \gamma)$. Hence, for $\gamma=1/2\pi= 0.16$ the growth time equals the rotation time.

The growth rates in the TI domain of Fig. \ref{map2} are given in Fig. \ref{growth1}. They are much larger than 0.16 so that the growth time of the 
instability is smaller than the rotation period. Note the rotational quenching of the  growth rates. As they are normalized with the inner 
rotation frequency, however, this is a formal effect. It means in Fig. \ref{growth1} that the physical growth time (in units of the dissipation 
time) does not depend on the rotation rate but it grows linearly with the magnetic field (see Goosens, Biront \& Tayler 1981).  

Figure \ref{growth1} (bottom)  provides the comparison of the growth rates of TI with those of AMRI. The values are calculated along the line  $\Re=500$ of Fig. 
\ref{map2}. While the growth time of the AMRI remains slower than the rotation period the TI modes become faster and faster for growing magnetic amplitudes.

\subsection{Magnetic limits}
Figure \ref{map4}  shows the limit of the subAlfv\'enic Tayler instability for rigid rotation but for various Pm in the  Rem-Ha plane. One finds the Pm-dependence of the resulting magnetic Mach numbers as rather weak. For $\Pm>1$ the Rem  are largest, 
i.e. for large Pm the rotational quenching is weak. It is stronger for $\Pm<1$ but it is strongest for $\Pm=1$ which is the favored application for all kinds of numerical simulations.

The  stabilization of the toroidal magnetic fields against Tayler instability in rigidly rotating radiative stellar zones is now considered. 
From Fig. \ref{map4} we take the simple relation 
\beg
 \rm Mm \simeq 0.5.
\label{12}
\ende
Large   Mm  means weak rotational quenching of the instability which appears to be realized for $\Pm\neq 1$.  
If (\ref{12}) is true  then
\beg
 B_\phi \gsim 2 \sqrt{\mu_0\rho} \ u^*
\label{13}
\ende
leads to instability, i.e. 
\beg
 B_\phi\, [{\rm Gauss}]  > 2 \ u^*\, [{\rm cm/s}]
\label{cond}
\ende 
when the linear rotation velocity $u^*$ of the stellar interior is taken in cm/s. One thus needs more than 100 kGauss 
to get a marginal TI instability of toroidal fields in  the rotating Sun. For Ap stars with their higher rotation velocities and higher 
mean densities even amplitudes of more than 10$^6$ Gauss are necessary. Due to the stellar rotation large-scale toroidal fields remain 
stable up to this value. Note that \citet{heger} work with toroidal field amplitudes of $10^4$ Gauss. For hot MS stars with their extended 
radiative zones and a linear rotation rate of (say) 100 km/s the minimum magnetic amplitude for the Tayler instability is $10^7$ Gauss. This result only concerns to the fast Tayler instability with growth times shorter than the rotation period. By the action of high thermal diffusion in the radiative  stellar cores also weaker toroidal fields can become unstable but with much larger growth times (see Fig.~\ref{onebelt}). Note that probably almost all rotating objects belong to the fast-rotating class with ${\rm Mm}>1$. For the exceptional case of galaxies   whose inner toroidal fields can be observed, the relation ${\rm Mm}>1$ is true. 

More calculations with smaller magnetic Prandtl numbers are needed to find the final results for the limits of the Tayler instability of 
toroidal fields in rapidly rotating stars. Of course, we can only speculate  whether the obtained  linear scaling also holds for very 
fast rotation.  Already small deviations from the linear dependence in (\ref{12}) would provide strongly different  magnetic 
field amplitudes. On the other hand, the present results from the global theory of magnetic 
instability of toroidal fields in regions of high thermal diffusion find the rotational stabilization of TI as very effective.  Because 
of the much weaker rotational quenching seen in Fig. \ref{map2} for the AMRI this sort of magnetic instability seems to be more important 
for stellar physics. We shall see that indeed this instability transports much more angular momentum than the TI. Note that the AMRI only 
exists under the presence of differential rotation.

Opposite to stars for galaxies we can measure their interior magnetic fields so that the magnetic Mach number  is well-known. It is  much larger than unity so that galaxies should be TI stable. However, because of their very high magnetic Prandtl number together with the particular galactic rotation law new calculations are needed in order to probe their stability.

\section{The angular momentum transport}
\citet{gelrued08} have shown that the angular momentum transport anticorrelates with $\nabla \Om$. A formulation as 
in (\ref{T1}) (`diffusion approximation') is indeed possible. One can  show that the quantity $T_R$ vanishes for rigid rotation and it is 
{\em linear} in $\nabla \Om$. It makes thus sense to define an eddy viscosity with a sofar unknown amplitude. In a linear theory a 
determination of the numerical value of the eddy viscosity  is not possible. We thus start 
with nonlinear simulations for a container with $\hat\eta=0.5$, also   to minimize the influence of the boundary conditions.

For the same reason the cylinder is regarded  with a periodicity in the vertical coordinate $z$ set to $6(\Rout-\Rin)$. We use the MHD Fourier 
spectral element code described by \citet{fournier} and \citet{gellert}. In this approach the solution is expanded in 
$M$ Fourier modes in the azimuthal direction. The resulting decomposition is a collection of meridional problems, each of which is
solved using a Legendre spectral element method (see e.g. \citealt{dev}). Either $M=8$ or $M=16$ Fourier modes are used, two or
three elements in radius and twelve or eighteen elements in axial direction, resp. The polynomial order is varied between $N=8$ 
and $N=16$. With a semi-implicit approach consisting of second-order backward differentiation formula and third order Adams-Bashforth for 
the nonlinear forcing terms time stepping is done with second-order accuracy. At the inner and outer wall no-slip and
perfect conducting boundary conditions are applied.

The Taylor-Couette flow (\ref{1}) as initial flow and white noise for the initial magnetic field are used. Additionally applied is the 
toroidal field with $\mu_B=1$ as external field. The differential rotation is varied between $\mu_\Omega=0.3$ and $\mu_\Omega=1.5$. 
At these parameter values  the field becomes unstable. In the linear regime only the Fourier mode $m=1$ grows 
exponentially. Already after roughly one rotation, when nonlinear effects become important, also higher modes appear. Though, the mode spectrum drops very 
quickly with increasing $m$. More than $99\%$ of the energy is contained within the first three modes. After only a few rotations all excited 
modes saturate and a steady-state is reached. The resulting fields are averaged in azimuthal direction and the nonaxisymmetric components are regarded 
as fluctuating quantities. 

\begin{figure}
\includegraphics[height=5cm,width=8cm]{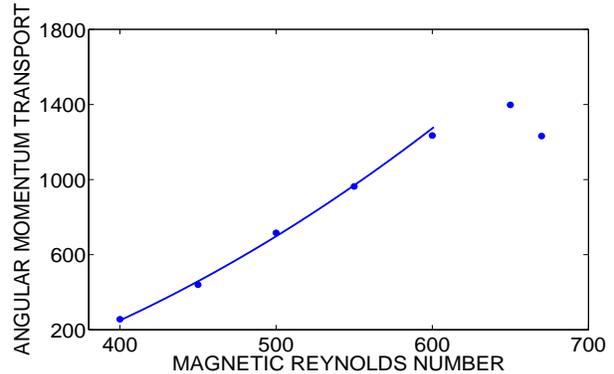}
\caption{Values of $T_R$ (in the center of the gap) for superAlfv\'enic rotation for $\etah=0.5, \ \mu_\Omega=0.5,\ \mu_B=1,\ \Pm=1$ and for 
         fixed $\Ha=250$. The fitted parabola shows $T_R$ depending on $\Rm^2$. For too high $\Rm$ the instability is suppressed so that always a maximum exists for fixed Ha at a certain rotation  rate. }
\label{fig_t_re}
\end{figure}

In accordance to Eq. (\ref{T}) the angular momentum transport is calculated. Figure \ref{fig_t_re} shows for subrotation ($\mu_\Omega=0.5$, 
quasigalactic rotation law), $\mu_B=1, \Pm=1$ and a Hartmann number of $\Ha=250$ how $T_R$ depends on the magnetic Reynolds number. 
For $T_R$ its values in the center of the gap and averaged over $z$ are used. The angular momentum transport scales with $\Rm^2$ up 
to $\Rm=600$. The last two points do not follow this rule because  the  upper  limit for the instability is approached. The 
differential rotation becomes too strong and suppresses the nonaxisymmetric magnetic instability equivalent to the results of   the linear 
analysis (Fig. \ref{map2}). Maxima as shown  in Fig. \ref{fig_t_re} basically exist due to the  limit  of  fast  rotation which always exists for the nonaxisymmetric magnetic instability. In order to find the  angular momentum transport of kink-type instabilities it is  necessary to find this maximum and to compute the maxima for increasing magnetic fields (see Fig. \ref{rapid2}, below). The question is how the maxima scale with increasing Ha. If they saturate for large Ha the characteristic angular momentum transport is known.

All the nonlinear simulations demonstrate that the angular momentum transport $T_R$ scales linearly with ${\rm d}\Om/{\rm d} R$ so that a characteristic eddy viscosity can be introduced.
In the following, these eddy viscosities are calculated in the two instability domains shown in Fig. \ref{map2}. The domain TI is 
characterized by subAlfv\'enic rotation and strong fields while in the  AMRI domain the rotation is superAlfv\'enic. We shall find that AMRI produces much 
higher values of eddy viscosity rather than TI.

\subsection{Slow rotation}
\begin{figure}
\includegraphics[height=4.1cm,width=8cm]{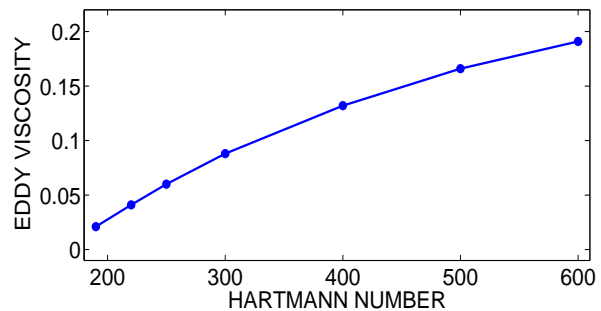}
\caption{Viscosities $\nu_{\rm T}/\nu$  for subAlfv\'enic rotation and  for $\mu_\Omega=0.5$. It is $\mu_B=1$, $\Re=30$, $\etah=0.5$, $\Pm=1$.}
\label{slow}
\end{figure}

We start to consider  the instability domain of   subAlfv\'enic rotation. In the  TI area of Fig. \ref{map2} for a small Reynolds number ($\Re=30$) and 
for various Hartmann numbers exceeding $\Ha=180$ the torque $T_R$  and the eddy viscosity are calculated. The eddy viscosity is normalized with the 
molecular viscosity so that the results are fully general. The normalized eddy viscosities prove to be very small. Note the  saturation for increasing 
Hartmann numbers (Fig. \ref{slow}).  We found the small values of the eddy viscosity as characteristic for the considered combination of strong fields and slow rotation. 

\subsection{Rapid rotation}
\begin{figure}
\includegraphics[height=5cm,width=8cm]{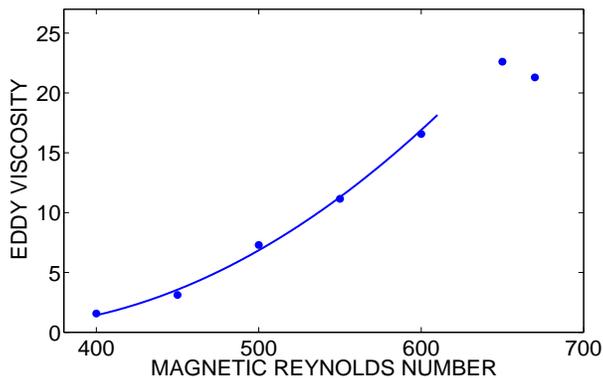}
\caption{The same as in Fig. \ref{fig_t_re} but for  $\nu_{\rm T}/\nu$. Note the existence of the maximum value due to  the rotational stabilization.}
\label{rapid1}
\end{figure}

\begin{figure}
\includegraphics[height=5cm,width=8cm]{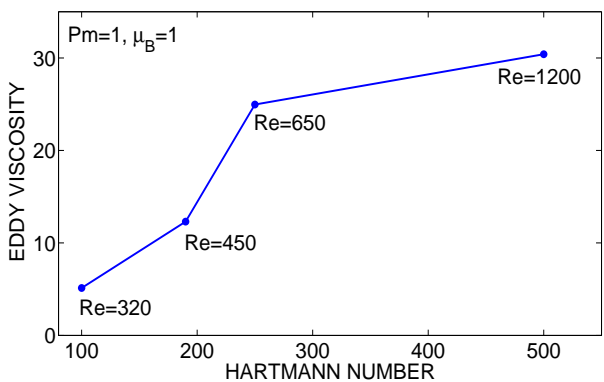}
\caption{The maximum $\nu_{\rm T}/\nu$ for various Hartmann numbers, $\Pm=1$. The curves are marked with the Reynolds number where the eddy 
         viscosity is largest. The normalized eddy viscosity seems to saturate for strong magnetic fields. Note that always  $\rm Mm>1$.}
\label{rapid2}
\end{figure}

The eddy viscosity is now calculated as a function of the Reynolds number of rotation for the same set of parameters used in Fig.~\ref{fig_t_re}. 
All these models belong to the domain AMRI in the instability map of Fig. \ref{map2}. The results are plotted in Fig. \ref{rapid1}. For $\nu_{\rm T}/\nu$ a 
scaling with $\Rm^2$ is  found in opposition to the $\Rm^{-1}$ scaling by \citet{maed05}. Due to the rotational quenching  of the instability    
a maximum exists   characteristic for the given magnetic field. The resulting viscosities $\nu_{\rm T}/\nu$ are much higher than for  subAlfv\'enic rotation but they  do 
not exceed the value of (say) 25. Figure  \ref{rapid2} shows the results of  calculations of maximum values for $\nu_{\rm T}/\nu$ for increasing magnetic amplitudes.  Also the Reynolds numbers 
belonging  to the maximum viscosity are given. We find  a saturation of the  ratio  $\nu_{\rm T}/\nu$  for strong magnetic fields. The eddy viscosity 
exceeds the microscopic viscosity not more than by a factor of 100. The same procedure must be applied to the other possible magnetic profiles in order to confirm the existence of the magnetic saturation suggested by Fig. \ref{rapid2}. The existence of the maximum of  $\nu_{\rm T}/\nu$ for fixed magnetic field as shown in Fig. \ref{rapid1} seems to be clear. The existence of the saturation for stronger magnetic fields must  be checked by further calculations.


\begin{figure}
\vbox{
\includegraphics[width=8cm]{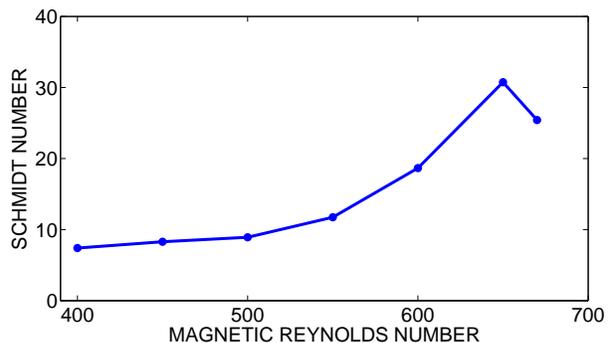}}
\caption{The Schmidt number ${\rm Sc}=\nu_{\rm T}/D_{\rm T}$ vs. Reynolds number Rm for $\Ha=250$ and $\Pm=1$.}
\label{fig_pd}
\end{figure}

We have also computed the expression (\ref{D}) for  turbulent diffusion with growth rates $\gamma$ taken from the initial growth phase of the instability. 
which unfortunately is a very simple approximation. Therefore, the  results do not have the same accuracy as the viscosity values. 
Values are approximately one order of magnitude below the eddy viscosity leading to high Schmidt
numbers ${\rm Sc} \approx 30$ (Fig. \ref{fig_pd}). We do not find a strong dependence on the magnetic field strength. It seems to be 
true that the kink-type instability for toroidal fields produces medium viscosity values by Maxwell stress and small diffusivity values by 
Reynolds stress. So far we cannot find a remarkable influence of the amplitude of the toroidal field on the resulting Schmidt number.

\section{Main results}

The stability of  toroidal magnetic fields is considered in cylindric geometry. The toroidal fields can be imagined as resulting from  the interaction of  weak poloidal fields with differential rotation. For $\Rm\gg 1$ it makes  sense to consider the stability for 
the MHD  system  under neglect of the poloidal field. In the present paper the stability/instability  of stationary toroidal fields against nonaxisymmetric perturbations with $m=1$ under the presence of rigid or differential rotation is discussed.

We have shown that under the influence of a rigid rotation the marginal instability of toroidal fields is basically suppressed. 
Note that the suppression is strongest for $\Pm=1$ and becomes weaker for any other value of Pm. For differential rotation the situation is more 
complicated. For superrotation (${\rm d} \Om/{\rm d} R>0$) the suppression is even stronger than for rigid rotation, while for subrotation 
(${\rm d} \Om/{\rm d} R<0$) it also  exists but only for  very fast rotation. These results are understandable because too strong differential 
rotation always disturbs nonaxisymmetric magnetic configurations. 

For radiative zones the kink-type  instability can be considered   as an angular momentum transporter. The angular 
momentum transport vanishes for solid-body rotation and  the angular momentum  is always transported opposite to the gradient of the angular velocity so that a `diffusion approximation´ with an eddy diffusivity makes sense.  The  eddy viscosity resulting from our simulations 
exceeds the microscopic viscosity not more than by 1--2 orders of magnitudes. The truth of this statement bases on the existence of the magnetic saturation of the maximum viscosity values presented in Fig. \ref{rapid2}.  Further simulations are needed to confirm this finding.

If angular momentum is transported and the transport is not completely magnetic then also 
the considered fluid  is continously mixed. For strong mixing  the stellar evolution would too strongly be affected (see \citealt{brott}), 
hence the  mixing induced by magnetic instability  {\em must} remain weak. Also the lithium observations of solar-type stars are not  understandable  without  the existence 
of  a rather mild turbulence  beneath the stellar convection zones. One may argue that the mixing is due to the Reynolds stress while  
the angular momentum transport is mainly due to the Maxwell stress. If the Maxwell stress dominates the Reynolds stress then the general transport problem  formulated 
by \cite{zahn} is solved. We  indeed  find such a  tendency  with nonlinear simulations. The eddy viscosity exceeds the diffusion coefficient by one 
or two orders of magnitude. Very similar results are reported by \citet{carballido} and \citet{Klahr} for simulations of the magnetorotational 
instability. 

It is an open question whether the dominance of eddy viscosity over turbulent diffusion can even be stronger. Figure~\ref{fig_pd} does not lead in 
this direction. More simulations with different magnetic profiles and/or with smaller magnetic Prandtl numbers must show how many orders the 
magnetic-induced angular momentum transport   exceeds the  diffusion of chemicals. 

Note that the values derived with our model -- in particular the value for the diffusion coefficient (\ref{D}) -- are maximum values as the 
suppressing action of the density stratification and the destabilizing action of the microscopic thermal diffusion is still neglected. Also 
the restriction to magnetic Prandtl number of order unity and the consideration of a special magnetic profile are still serious limitations. 

\label{lastpage}



\end{document}